# Fabrication of Nb/Al₂O₃/Nb Josephson Junctions using *in situ* Magnetron Sputtering and Atomic Layer Deposition

Rongtao Lu, Alan J. Elliot, Logan Wille, Bo Mao, Siyuan Han, Judy Z. Wu, John Talvacchio, Heidi M. Schulze, Rupert M. Lewis, Daniel J. Ewing, H. F. Yu, G. M. Xue, and S. P. Zhao

*Abstract*— **Atomic layer deposition (ALD) provides a promising approach for deposition of ultrathin low-defect-density tunnel barriers, and it has been implemented in a high-vacuum magnetron sputtering system for *in situ* deposition of ALD-Al₂O₃ tunnel barriers in superconductor-insulator-superconductor (SIS) Josephson junctions. A smooth ALD-Al₂O₃ barrier layer was grown on a Al-wetted Nb bottom electrode and was followed with a top Nb electrode growth using sputtering. Preliminary low temperature measurements of current-voltage characteristics (IVC) of the Josephson junctions made from these trilayers confirmed the integrity of the ALD-Al₂O₃ barrier layer. However, the $I_cR_N$ product of the junctions is much smaller than the value expected from the Ambegaokar-Baratoff formula suggesting a significant pair-breaking mechanism at the interfaces.**

## I. INTRODUCTION

Josephson junctions are the central element of superconducting quantum bits (qubits) which is one of the most promising approaches to realizing scalable quantum computing. While tremendous progress has been made in research and development of superconducting qubits [1], further enhancement of the quantum coherence time is necessary. It has been reported that defects in superconducting junctions, especially in the dielectric tunnel barrier layer, acting as two-level systems (TLS) which has become the main source of decoherence in junctions [2]. A recent work by Oh *et al.* revealed that the density of the TLS defects could be

Manuscript received October 9, 2012. This work was supported in part by the ARO grand W911NF-09-1-0295 and W911NF-12-1-0412. JW also acknowledges support from NSF contracts NSF-DMR-1105986 and NSF EPSCoR-0903806, and matching support from the State of Kansas through Kansas Technology Enterprise Corporation. The work in Northrop Grumman was supported in part by DMEA contract H94003-04-D-0004-0149. The work at Institute of Physics, Chinese Academy of Sciences was supported by the National Natural Science Foundation of China (Grant No.11104340) and 973 Program (Grants No. 2009CB929102).

R.T. Lu, A.J. Elliot, L. Willie, B. Mao, S.Y. Han, and J.Z. Wu are with the Department of Physics and Astronomy, University of Kansas, Lawrence, Kansas 66045, USA (corresponding author: R.T. Lu, 785-864-2274, fax: 785-864-5262, e-mail: rtlu@ku.edu; J.Z. Wu, 785-864-3240, fax: 785-864-5262, email: jwu@ku.edu).

J. Talvacchio, H.M. Schulze, R.M. Lewis, and D.J. Ewing are with Northrop Grumman, Baltimore, MD 21203, USA (e-mail: john.talvacchio@ngc.com).

H.F. Yu, G.M. Xue and S.P. Zhao are with Beijing National Laboratory for Condensed Matter Physics, Institute of Physics, Chinese Academy of Sciences, Beijing 100190, China (e-mail: spzhao@iphy.ac.cn).

significantly reduced by replacing amorphous $AlO_x$ barrier with epitaxial $Al_2O_3$ layer on superconducting Re electrode [3, 4], which led to significant improvement of coherence time. However, the required high temperature post-annealing at temperatures in the neighborhood of 800 ºC seems necessary for crystallization of the $Al_2O_3$ tunnel barrier layer, which induces rough interfaces between the superconductors and the tunnel barrier. In particular, the high mobility of Al at such high processing temperatures prevents *in situ* epitaxial growth of the superconductor-insulator-superconductor (SIS) trilayers in a layer-by-layer fashion [3]. Meanwhile, for the two most widely used material systems of superconducting qubits, Nb-$AlO_x$-Nb and Al-$AlO_x$-Al, epitaxial growth of $Al_2O_3$ will be extremely challenging due to the fundamental difference in lattice structures between the hexagonal $Al_2O_3$ and the cubic Nb and Al. Therefore, exploration of new approaches for fabrication of ultrathin (on the order of 1 nm) tunnel barrier with greatly reduced defect density is imperative to increasing the performance of superconducting qubits.

Atomic layer deposition (ALD) of insulating tunnel barriers may provide a promising solution to this long standing problem. ALD has multiple advantages including conformal growth, atomic-scale thickness control, and low defect density. Such an accurate growth control is attributed to the self-limited mechanism of chemical adsorption and reaction in ALD [5], and it is important to obtain a tunnel barrier layer with the required small thickness and low defect density. Especially, the self-limited adsorption of source vapor and sequential chemical reaction in the ALD process most likely result in a complete layer-by-layer oxidation, which differs fundamentally from the physical diffusion in thermal oxidation process and could lead to much lower defect densities. Thus ALD technique is very well suited for fabricating ultrathin, low defect density tunnel barrier in SIS junctions. In an earlier attempt using a "pseudo-atomic layer deposition" by repeatedly applying ultra thin film sputtering/oxidation, fabrication of tunnel barriers in the layer-by-layer fashion was reported [6]. While this differs from the ALD approach reported in this work, the two methods share the motivation in replacement of oxygen diffusion with layer-by-layer fabrication of the tunnel barrier.

ALD growths of $Al_2O_3$ films on semiconducting and insulating substrates have been widely explored during the past two decades [7]. However, the ALD-$Al_2O_3$ growth requires hydroxyl bonding on the sample surface, which means direct ALD-Al$_2$O$_3$ growth on metal surfaces is challenging due to the lack of a nucleation mechanism such as hydroxyl groups [8, 9]. To initiate ALD nucleation on metals, one approach is to introduce hydroxylation, and this may be accomplished using a hydrous plasma [9]. The hydroxylation may also be obtained naturally on some metals that can be easily oxidized by a similar process of hydroxylation, which happens on semiconductors such as Si via attachment of hydroxyl group to the surface oxygen. Nb is a well known candidate in terms of its ease of oxidation. However, its multiple valence states could produce multiple NbO$_x$ species with electrical properties that range from metallic, to semiconducting, to insulating [10]. This also typically leads to a nonuniform surface oxide layer, which is not suitable for Josephson tunnel barriers. Consequently, direct oxidation of Nb needs to be avoided. Al is a promising candidate as the wetting layer for Nb since it has high wettability and ease of oxidation [11, 12]. In particular, thermal AlO$_x$ has been used as the tunnel barrier for Nb/AlO$_x$/Nb Josephson tunnel junctions. To minimize the contribution from naturally formed AlO$_x$ to the tunneling properties of the Josephson junctions, precautions were taken to minimize exposing the Al wetting layer's surface to oxygen. One major issue in fabricating SIS trilayers using ALD stems from the difficulties in interfacing between the commercial ALD chamber with high-vacuum (HV) or ultra-high-vacuum (UHV) physical vapor deposition (PVD) systems, such as magnetron sputtering, for in-situ fabrication of multilayer films. In order to resolve this issue, a home designed ALD system was assembled and interfaced with an UHV sputtering system. In this work, we report the fabrication of Nb/ALD-Al$_2$O$_3$/Nb trilayers with Al wetting layers using this in-situ sputtering/ALD system. Preliminary characterizations have demonstrated the integrity of leak-free ultrathin tunnel barriers made with Nb/ALD-Al$_2$O$_3$/Nb Josephson junctions. The range of critical current density $J_c$ for superconducting qubits is between 1 A/cm$^2$ to 10$^2$ A/cm$^2$, depending on qubit designs. Normally, this corresponds to a specific resistance $R_NA$ in the range of ∼ 220 kΩ·μm$^2$ to 2.2 kΩ·μm$^2$ for Nb junctions and ∼ 25 kΩ·μm$^2$ to 250 Ω·μm$^2$ for Al junctions. In our experiments, a normal state resistance of 220 Ω and a specific resistance of 3.57 kΩ·μm$^2$ have been obtained in the sample with 8 cycles of ALD-Al$_2$O$_3$, and a sub-gap resistance ∼ 4.5 kΩ and sub-gap/normal-state resistance ratio of 20.5 have been obtained.

## II. EXPERIMENT

Si substrates with a 500 nm thick thermal oxide layer were mounted on a water-cooled stage for Nb sputtering. 90-200 nm thick Nb films were fabricated using DC sputtering from a 3 inch Nb target in the main sputtering chamber at 14 mTorr Ar, 330W. The Nb film sputtering rate was calibrated by using a KLA Tencor P-16 profiler, which was also used to scan three-dimensional morphology of the patterned junctions. After the deposition of the bottom Nb films, a 7 nm thick Al layer was deposited using DC sputtering from a 3 inch Al target at 14 mtorr and 90W as a wetting layer for ALD-Al$_2$O$_3$ growth. The samples were then transferred into an ALD chamber connected to the sputtering chamber without breaking vacuum for in-situ ALD-Al$_2$O$_3$ growth.

The formula of fabricating ALD-Al$_2$O$_3$ [7] using trimethylaluminum (TMA, Al(CH$_3$)$_3$; semiconductor grade, Akzo Nobel) and water (optima grade, Fisher Scientific) is shown in Fig. 1(a). The ALD-Al$_2$O$_3$ process is schematically described in Fig. 1(b). Ultrahigh purity (99.999%) N$_2$ was used as carrier gas to carry source vapor into the ALD chamber and also worked as a purge gas. A water pulse was applied at the beginning and the hydroxyl groups were anticipated to bond on the Al wetting layer surface. This pretreatment of the Al surface was followed with the standard ALD cycles, each of which consisted of four steps: TMA exposure, purge with N$_2$, water exposure, and purge with N$_2$. The TMA exposure allows the Al-methyl group to be adsorbed and bonded atop of the hydroxyl group. The N$_2$ purge removes residual TMA after the full surface coverage of TMA to prepare for the water molecule absorption and hydroxyl group bond atop. The cycle is completed with the second N$_2$ purge before the next cycle begins.

The ALD chamber was pumped to 10$^{-6}$ Torr when the sample was transferred from the main sputtering chamber. Then the ALD chamber was heated up to 200 °C using resistive heaters, and chamber pressure was maintained at about 400-500 mTorr with a 5 sccm N$_2$ flow during the entire heating time of about 1.5 hrs. The ALD source vapor exposure and N$_2$ purge were achieved by using computer controlled solenoid valves. The exposure times for TMA and water pulses were both 5 seconds and the two purge times were 30 seconds. A quartz crystal monitor (QCM) was used for real-time growth monitoring. After the ALD was completed, the samples were cooled down naturally in about 1 hour and transferred back to the sputtering chamber for the top Nb layer sputtering. The diagram of the final trilayer is shown in Fig. 1(c). ALD-Al$_2$O$_3$ sample with 60 cycles grown on Al wetting layer was used for ALD-Al$_2$O$_3$ growth rate calibration using a Horiba UVISEL spectroscopic ellipsometer between 2.75 eV and 4 eV. The thick ALD barrier prevented further oxidation through Al layer after deposition and made growth rate calibration possible. Other control samples that only went through the ALD heating/cooling process without exposure to any ALD source vapor were also made for junction fabrication and low temperature characterization.

Fig. 2 shows the measured chamber pressure (top) and QCM resonate frequency (bottom) versus time in an 8 cycle ALD-Al$_2$O$_3$ fabrication process on top of Nb(200 nm)/Al(7 nm). The water and TMA pulses can be clearly identified from the figure with pulse heights in the range of 30-60 mTorr relative to carrier gas pressure. The observed QCM resonance frequency shows multiple steps, and each step exactly correspond to either a water or a TMA pulse, indicating the efficient adsorption of source vapors and growth of ALD-Al$_2$O$_3$ film. The calculated frequency change of about 10 Hz per cycle, together with the measured ALD-Al$_2$O$_3$ growth rate of about 0.12 nm/cycle using spectroscopic ellipsometer, can be used to monitor the growth of ALD-Al$_2$O$_3$ barrier.

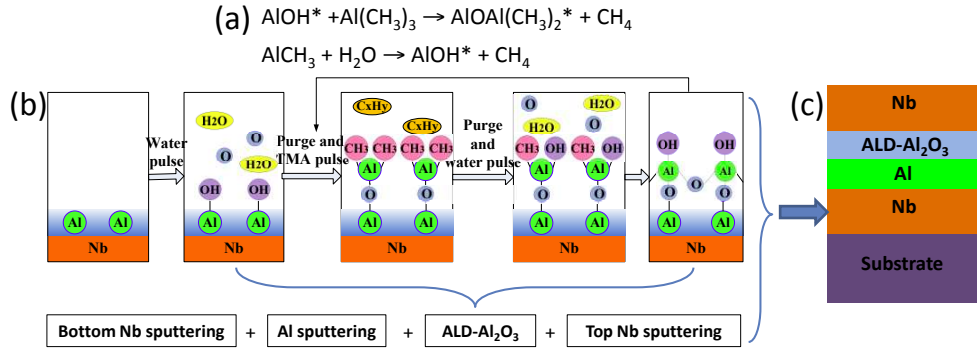

Fig. 1 ALD based trilayer fabrication process. (a) Reaction formula of ALD-Al$_2$O$_3$ using TMA and water. (b) Flow chart of ALD-Al$_2$O$_3$ fabrication. (c) Diagram of trilayer with ALD-Al$_2$O$_3$ barrier.

Resistance vs. temperature variation was measured using standard four-probe measurement. Surface morphology was characterized using atomic force microscopy (AFM). Room temperature current-in-plane tunneling (CIPT) measurement [13, 14] was used to diagnose the existence of ALD barrier layer by measuring the tunneling resistance at room temperature. Josephson junctions with 4 $\mu$m × 4 $\mu$m area were fabricated using this ALD-trilayer and the current - voltage curve (IVC) was characterized at liquid helium temperature [5, 15]. To minimize electromagnetic interference from environment, a trilayer $\mu$-metal magnetic shield is used to reduce the ambient static field to about 20 nT. In addition, all electrical leads used for measuring IVCs are carefully filtered by cryogenic RC low-pass filters and copper powder filters (CPF). The experimental setup is described in detail in Ref. [15].

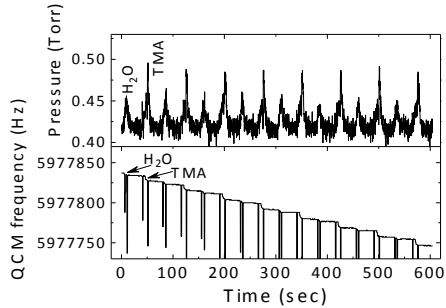

Fig. 2 Pressure of TMA/H$_2$O pulses (top) and the corresponding QCM frequency change (bottom) in an 8 cycles ALD-Al$_2$O$_3$ process.

## III. RESULTS AND DISCUSSIONS

The AFM images of the Al wetting layer and the ALD-Al$_2$O$_3$ barrier are shown in Fig. 3. The surface of a 7 nm thick Al wetting layer on top of 200 nm thick Nb film is smooth with an average roughness $R_a$ of 0.87 nm [Fig. 3(a)]. After 14 cycles of ALD-Al$_2$O$_3$ growth, the surface was still smooth with a roughness of $R_a \sim 0.63$ nm [Fig. 3(b)], which is comparable to that of Al wetting layer and may be attributed to the conformal growth of ALD. The smooth surface of ALD-Al$_2$O$_3$ will further benefit the fabrication of high quality trilayers and tunnel junctions.

To characterize the tunnel properties of the trilayers, CIPT measurement was employed to test and verify the existence of an ALD barrier as a convenient, room temperature method. The CIPT technique was initially used to measure the magneto-resistance of magnetic tunnel junctions, and the tunneling resistance in multi-layer films can be estimated using CIPT without patterning the sample [13, 14]. For a Nb(92 nm)/Al(7 nm)/Nb(92 nm) reference trilayer that only went through the ALD heating/cooling processes without exposure to any source vapor, the tunneling resistance was too low to measure using CIPT, indicating the heating/cooling process in ALD does not cause significant oxidation of the Al wetting layer. The result is expected since the saturation thickness of an oxide film on aluminum can be tuned by controlling oxygen pressure [16], indicating very thin AlO$_x$ might have been formed on Al wetting layer during the ALD heating/cooling processes, corresponding to a very high $J_c$ of 9.5 kA/cm$^2$ observed on the reference sample, as shown in Fig. 4(a). For a Nb(92 nm)/Al(7 nm)/ALD-Al$_2$O$_3$(14 cycles)/Nb(92 nm) trilayer with 14 ALD cycles grown on top of the Al wetting layer, the tunneling resistance was clearly identified by CIPT measurement, indicative of the ALD-Al$_2$O$_3$ tunnel barrier formation in ALD growth. In fact, we have varied the number of the ALD cycles in the range of 2-20, and a monotonic increase trend of the tunneling resistance with the number of the ALD cycles has been observed. In addition, uniform tunneling resistance with a small standard deviation of less than 10% was observed on most samples with diameters up to 50 mm.

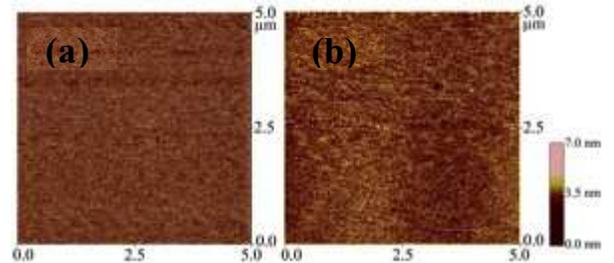

Fig. 3 AFM images of (a) Al(7nm)/Nb(200nm), $R_a$ = 0.87 nm; (b) 14 cycles of ALD-Al$_2$O$_3$ on top of Nb(200nm)/Al(7nm), $R_a$=0.63 nm. Scan area is 5 $\mu$m × 5 $\mu$m.

Direct low temperature characterization of tunneling properties was attempted on the Josephson junctions fabricated from ALD trilayers, and the data obtained from representative trilayers are depicted in Fig. 4. The $R$-$T$ curve of a 200 nm thick Nb film shows a transition temperature of about 9.0 K and a resistance ratio $R_{RT}/R_{10K} \sim$3.0, which are expected for good quality Nb films. Comparable $R_{RT}/R_{10K} \sim$ 3.0 was also obtained from the top Nb electrode of an ALD trilayer, indicating the top Nb film grown on the ALD-Al$_2$O$_3$ layer has similar quality. Josephson junctions were fabricated using some of the ALD trilayers. For a 4 μm × 4 μm junction made from the reference trilayer which only went through ALD heating/cooling process but without exposure to ALD sources, its IVC at 4.2K [Fig. 4(a)] shows clearly a sudden increase of tunneling current at approximately twice of the niobium's superconducting gap voltage $2\Delta/e \sim 2.2$ mV, where $\Delta$ and $e$ are the superconducting gap energy and the charge of an electron, respectively. The specific resistance of this junction is $R_N A \sim 14.7$ $\Omega\mu m^2$, where $R_N$ and $A$ are the normal resistance and area of the junction, respectively. Assuming $I_c R_N$ product is proportional to $\Delta$ [17], the critical current $I_c$ of the reference junction can be estimated using $e I_c R_N/\Delta = 1.27$, which is determined from a large number of Nb/AlO$_x$/Nb junctions previously measured in our lab [18]. We hence obtain $I_c R_N \sim 1.4$ mV for the reference junction. In the absence of extra pair-breaking mechanisms, the critical current density $J_c$ of the Josephson junctions made from this reference trilayer is expected to be $J_c \sim 9.5$ kA/cm$^2$ at 4.2K. The expected high $J_c$ value indicates a very thin tunnel barrier that was most likely formed on the top surface by thermal oxidation of the Al wetting layer during the heating/cooling process prior to the introduction of ALD sources into the ALD chamber. Fig 4(b) shows the IVC of a 4 μm × 4 μm junction fabricated from an ALD trilayer of Nb(150 nm)/ALD-Al$_2$O$_3$(8 cycles)/Nb(50 nm) with Al wetting layer. The quasiparticle branch of the IVC has a very low sub-gap leakage current and the much higher specific resistance of $\sim 3.57$ k$\Omega\mu m^2$ shows that the tunnel barrier layer is dominated by ALD-grown Al$_2$O$_3$ instead of the thermal oxidation of the Al wetting layer. The sub-gap resistance $R_{sg} \sim 4.5$ k$\Omega$ was measured at $V=\pi\Delta/2e \sim 1.8$ mV, and the sub-gap/normal-state resistance ratio $R_{sg}/R_N \approx 20.5$ is comparable to previous reports using AlN tunneling barriers [19, 20]. However, $I_c R_N$ product of the junctions made from the reference trilayer and the ALD trilayer are only about 0.3 mV to 0.5 mV which is a factor of three to five less than that expected from the Ambegaokar-Baratoff formula and/or empirical results compiled from junctions made from Nb/Al-AlO$_x$/Nb trilayers with controlled thermal oxidation. Since both, the reference and the ALD trilayer, have the same problem of significantly suppressed $J_c$ values, they seem to suffer from the same pair-breaking mechanism. In principle, magnetic impurities and charge trap centers could result in significantly reduced pair current. However, in our experiment the probability of having contamination from magnetic impurities is extremely low leaving charge scattering centers as the most probable source of extra pair breaking. We suspect that for the ALD-Al$_2$O$_3$ barrier, the hydroxyl group terminated surface might originate scattering centers. Moreover, the ALD process was performed in a low vacuum environment with nitrogen carrier gas and a pressure of a few $10^2$ mTorr was maintained using a mechanical pump, which could introduce a little thermal oxidation or defects into the ALD tunnel barrier. Such a low vacuum background was also maintained by using mechanical pumping during heating/cooling process and makes it hard to remove the water vapor residue. Thus the residual hydroxyl adsorbed on top of Al wetting layer may introduce charge scatter centers in the control sample shown in Fig. 4(a). Upgrading to high-vacuum based ALD fabrication may reduce the possibility of introducing additional hydroxyl groups into the sample prior to ALD growth, and further optimizations on the trilayer fabrication and junction performances are ongoing. It should be realized that other pair-breaking mechanisms may also exist. For example, diffusion of H, C, N, O radicals or molecules through the tunnel barrier/electrode interfaces into Al or/and Nb electrode may also be important. Further characterization of the ALD tunnel barriers is necessary to pinpoint the pair-breaking mechanisms and to develop schemes to eliminate them.

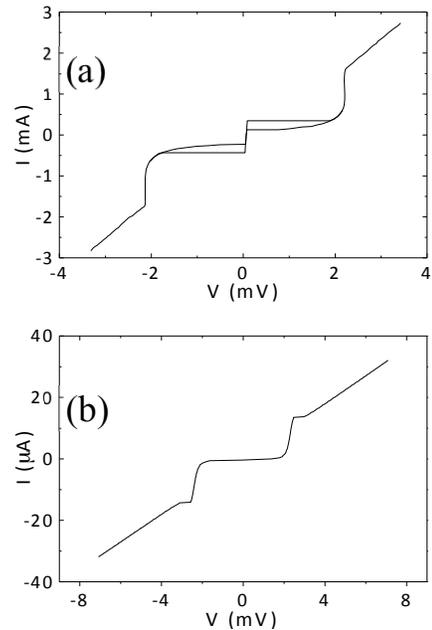

Fig. 4 (a) IVC of a 4 μm × 4 μm Josephson junction made from the Nb(150 nm)/Al(7 nm)/Nb(50 nm) reference trilayer which only went through heating/cooling process without exposure to any ALD source. The IVC was measured using a two-probe method. (b) IVC of a 4 μm × 4 μm Josephson junction made from the ALD trilayer of Nb(150 nm)/ALD-Al$_2$O$_3$(8 cycles)/Nb(50 nm) with a 7-nm Al wetting layer. The junction was measured using the four-probe method. All IVC measurements were made at 4.2K.

## IV. CONCLUSION

An integrated magnetron sputtering/ALD system was assembled for *in-situ* fabrication of the Al$_2$O$_3$ tunnel barriers in the Nb/ALD-Al$_2$O$_3$/Nb SIS trilayers. An Al wetting layer deposited on Nb base electrode has been adopted to assist the adsorption of ALD source vapors. Growth of ALD barrier was monitored using a QCM monitor. The formation of ALD-Al$_2$O$_3$ tunnel barrier has been confirmed by room temperature CIPT measurement on unpatterned trilayers. Further definitive proofs of the tunneling behaviors have been observed in the low temperature IVC on ALD Josephson tunnel junctions. The observed $I_c$ suppression was attributed to pair-breaking induced by hydroxyl scattering centers. Further optimizations are ongoing to improve the junction performance and explore more fundamental details for the ALD growth process of the SIS trilayer.